\RequirePackage{lineno}
\documentclass[aps,prl,twocolumn,floatfix,superscriptaddress,showpacs]{revtex4-1}
\usepackage{graphicx}
\usepackage{amsmath}
\usepackage{dcolumn}
\usepackage{array}
\usepackage{amssymb}

\begin{document}
%\linenumbers

\title{Jet-Hadron Correlations in $\sqrt{s_{\text{NN}}} = 200~\text{GeV}$ $p$+$p$ and Central Au+Au Collisions}
\date{\today}
\affiliation{AGH University of Science and Technology, Cracow, Poland}
\affiliation{Argonne National Laboratory, Argonne, Illinois 60439, USA}
\affiliation{University of Birmingham, Birmingham, United Kingdom}
\affiliation{Brookhaven National Laboratory, Upton, New York 11973, USA}
\affiliation{University of California, Berkeley, California 94720, USA}
\affiliation{University of California, Davis, California 95616, USA}
\affiliation{University of California, Los Angeles, California 90095, USA}
\affiliation{Universidade Estadual de Campinas, Sao Paulo, Brazil}
\affiliation{Central China Normal University (HZNU), Wuhan 430079, China}
\affiliation{University of Illinois at Chicago, Chicago, Illinois 60607, USA}
\affiliation{Cracow University of Technology, Cracow, Poland}
\affiliation{Creighton University, Omaha, Nebraska 68178, USA}
\affiliation{Czech Technical University in Prague, FNSPE, Prague, 115 19, Czech Republic}
\affiliation{Nuclear Physics Institute AS CR, 250 68 \v{R}e\v{z}/Prague, Czech Republic}
\affiliation{Frankfurt Institute for Advanced Studies FIAS, Germany}
\affiliation{Institute of Physics, Bhubaneswar 751005, India}
\affiliation{Indian Institute of Technology, Mumbai, India}
\affiliation{Indiana University, Bloomington, Indiana 47408, USA}
\affiliation{Alikhanov Institute for Theoretical and Experimental Physics, Moscow, Russia}
\affiliation{University of Jammu, Jammu 180001, India}
\affiliation{Joint Institute for Nuclear Research, Dubna, 141 980, Russia}
\affiliation{Kent State University, Kent, Ohio 44242, USA}
\affiliation{University of Kentucky, Lexington, Kentucky, 40506-0055, USA}
\affiliation{Korea Institute of Science and Technology Information, Daejeon, Korea}
\affiliation{Institute of Modern Physics, Lanzhou, China}
\affiliation{Lawrence Berkeley National Laboratory, Berkeley, California 94720, USA}
\affiliation{Massachusetts Institute of Technology, Cambridge, Massachusetts 02139-4307, USA}
\affiliation{Max-Planck-Institut f\"ur Physik, Munich, Germany}
\affiliation{Michigan State University, East Lansing, Michigan 48824, USA}
\affiliation{Moscow Engineering Physics Institute, Moscow Russia}
\affiliation{National Institute of Science Education and Research, Bhubaneswar 751005, India}
\affiliation{Ohio State University, Columbus, Ohio 43210, USA}
\affiliation{Old Dominion University, Norfolk, Virginia 23529, USA}
\affiliation{Institute of Nuclear Physics PAN, Cracow, Poland}
\affiliation{Panjab University, Chandigarh 160014, India}
\affiliation{Pennsylvania State University, University Park, Pennsylvania 16802, USA}
\affiliation{Institute of High Energy Physics, Protvino, Russia}
\affiliation{Purdue University, West Lafayette, Indiana 47907, USA}
\affiliation{Pusan National University, Pusan, Republic of Korea}
\affiliation{University of Rajasthan, Jaipur 302004, India}
\affiliation{Rice University, Houston, Texas 77251, USA}
\affiliation{University of Science and Technology of China, Hefei 230026, China}
\affiliation{Shandong University, Jinan, Shandong 250100, China}
\affiliation{Shanghai Institute of Applied Physics, Shanghai 201800, China}
\affiliation{SUBATECH, Nantes, France}
\affiliation{Temple University, Philadelphia, Pennsylvania 19122, USA}
\affiliation{Texas A\&M University, College Station, Texas 77843, USA}
\affiliation{University of Texas, Austin, Texas 78712, USA}
\affiliation{University of Houston, Houston, Texas 77204, USA}
\affiliation{Tsinghua University, Beijing 100084, China}
\affiliation{United States Naval Academy, Annapolis, Maryland, 21402, USA}
\affiliation{Valparaiso University, Valparaiso, Indiana 46383, USA}
\affiliation{Variable Energy Cyclotron Centre, Kolkata 700064, India}
\affiliation{Warsaw University of Technology, Warsaw, Poland}
\affiliation{University of Washington, Seattle, Washington 98195, USA}
\affiliation{Wayne State University, Detroit, Michigan 48201, USA}
\affiliation{Yale University, New Haven, Connecticut 06520, USA}
\affiliation{University of Zagreb, Zagreb, HR-10002, Croatia}

\author{L.~Adamczyk}\affiliation{AGH University of Science and Technology, Cracow, Poland}
\author{J.~K.~Adkins}\affiliation{University of Kentucky, Lexington, Kentucky, 40506-0055, USA}
\author{G.~Agakishiev}\affiliation{Joint Institute for Nuclear Research, Dubna, 141 980, Russia}
\author{M.~M.~Aggarwal}\affiliation{Panjab University, Chandigarh 160014, India}
\author{Z.~Ahammed}\affiliation{Variable Energy Cyclotron Centre, Kolkata 700064, India}
\author{I.~Alekseev}\affiliation{Alikhanov Institute for Theoretical and Experimental Physics, Moscow, Russia}
\author{J.~Alford}\affiliation{Kent State University, Kent, Ohio 44242, USA}
\author{C.~D.~Anson}\affiliation{Ohio State University, Columbus, Ohio 43210, USA}
\author{A.~Aparin}\affiliation{Joint Institute for Nuclear Research, Dubna, 141 980, Russia}
\author{D.~Arkhipkin}\affiliation{Brookhaven National Laboratory, Upton, New York 11973, USA}
% ...

\collaboration{STAR Collaboration}\noaffiliation


\begin{abstract}
Azimuthal angular correlations of charged hadrons with respect to the axis of a reconstructed (trigger) jet in Au+Au and $p$+$p$ collisions at $\sqrt{s_{\text{NN}}} = 200~\text{GeV}$ in STAR are presented. The trigger jet population in Au+Au collisions is biased towards jets that have not interacted with the medium, allowing easier matching of jet energies between Au+Au and $p$+$p$ collisions while enhancing medium effects on the recoil jet. The associated hadron yield of the recoil jet is significantly suppressed at high transverse momentum ($p_{\text{T}}^{\text{assoc}}$) and enhanced at low $p_{\text{T}}^{\text{assoc}}$ in 0-20\% central Au+Au collisions compared to $p$+$p$ collisions, which is indicative of medium-induced parton energy loss in ultrarelativistic heavy-ion collisions.
\end{abstract}

\maketitle

High-energy collisions of heavy nuclei at the Relativistic Heavy Ion Collider (RHIC) at Brookhaven National Laboratory produce an energy density at which a strongly-coupled medium of deconfined quarks and gluons, known as the Quark-Gluon Plasma, is expected to form~\cite{STARwhite,PHENIXwhite,PHOBOSwhite,BRAHMSwhite}.  The properties of this medium can be probed using partons with large transverse momenta ($p_{\text{T}}$) resulting from hard scatterings in the initial stages of the collision.  The scattered partons recoil and fragment into back-to-back clusters of hadrons, known as jets.  

Jets in $p$+$p$ collisions are well-described by perturbative quantum chromodynamics (pQCD)~\cite{pQCD} and can be used as a reference for studies of medium-induced jet modification.  By comparing the jet momentum spectra as well as the momentum and angular distributions of jet fragments between heavy-ion collisions and elementary collisions, it is possible to investigate the energy loss of fast-moving partons in the QGP.  
%Theoretical models of partonic energy loss include radiative/collisional energy loss models~\cite{radTheoryFirst,radTheory0,radTheory1,radTheory2,radTheory3,radcollReview}, in which jets are softened and broadened due to interactions with the medium.  

Jet physics in heavy-ion collisions is frequently studied by using high-$p_{\text{T}}$ hadrons as jet proxies.  Suppression of high-$p_{\text{T}}$ hadrons in single-particle measurements, and of particle yields on the recoil side (``awayside'') of high-$p_{\text{T}}$ triggered ``dihadron'' correlations, has been observed at $\sqrt{s_{\text{NN}}} = 200~\text{GeV}$ at RHIC in central Au+Au collisions relative to $p$+$p$ and $d$+Au collisions~\cite{STARraa, PHENIXraa, STARdihadron0,STARdihadron, STARdihadron2, STARdihadron3, PHENIXdihadron, PHENIXdihadron2}, and at $\sqrt{s_{\text{NN}}} = 2.76~\text{TeV}$ at the Large Hadron Collider (LHC) in Pb+Pb collisions relative to $p$+$p$ and $p$+Pb collisions~\cite{ALICEraa, CMSraa, ALICErppb, ALICEdihadron, CMSdihadron}.  This suppression of jet fragments is often attributed to partonic energy loss due to interactions with the medium~\cite{radTheoryFirst}.  

In elementary collisions jets can be reconstructed by clustering their constituents in order to determine the energy and direction of the parent parton~\cite{jade,ktalgo1,ktalgo2}.  However, full jet reconstruction in a heavy-ion environment presents large challenges due to the fluctuating underlying event from soft processes.  Advancements in jet-finding techniques~\cite{fastjet2}, as well as the proliferation of high-$p_{\text{T}}$ jets at the energies accessible at the LHC, have made it possible to study fully-reconstructed jets in heavy-ion collisions for the first time.  Measurements of the dijet imbalance~\cite{ATLASdijet,CMSdijet}, fragmentation function~\cite{CMSfrag}, and jet $R_{AA}$ and $R_{CP}$~\cite{ATLASjetrcp}, among others, are being used to constrain models of jet quenching at LHC energies.  

At RHIC energies it is now possible to study triggered correlations with respect to the axis of a reconstructed jet, instead of using the dihadron correlation technique in which a high-$p_{\text{T}}$ hadron is used as a proxy for the jet axis.  Jet reconstruction allows more direct access to the original parton energy and makes it possible to select a sample of higher-energy partons, thus increasing the kinematic reach of these correlation measurements.  In this analysis, azimuthal angular correlations of mid-rapidity charged hadrons are studied with respect to a reconstructed mid-rapidity (trigger) jet.  The effects of medium-induced partonic energy loss, or ``jet-quenching,'' can be studied by comparing the shapes and associated hadron yields of jets in Au+Au with those in $p$+$p$ collisions.  

The data used in this analysis were collected by the STAR detector at RHIC for $p$+$p$ and Au+Au collisions at $\sqrt{s_{\text{NN}}} = 200~\text{GeV}$ in 2006 and 2007, respectively.  Charged tracks are reconstructed in the Time Projection Chamber (TPC)~\cite{STARtpc} and the transverse energy ($E_{\text{T}}$) of neutral hadrons is measured in the Barrel Electromagnetic Calorimeter (BEMC) towers (with azimuthal angle $\times$ pseudorapidity size $\Delta \phi \times \Delta \eta = 0.05 \times 0.05$)~\cite{STARbemc}.  Energy deposited by charged hadrons in the BEMC is accounted for by the hadronic correction, in which the transverse momentum of any charged track pointing towards a tower is subtracted from the transverse energy of that tower.  

Events are selected by an online high tower (HT) trigger, which requires $E_{\text{T}} \gtrsim 5.4~\text{GeV}$ in at least one BEMC tower.  An offline HT threshold of $E_{\text{T}} > 6~\text{GeV}$ is imposed (after hadronic correction).  In Au+Au only the 20\% most central events are analyzed, where event centrality is determined by the uncorrected charged particle multiplicity in the TPC within pseudorapidity $|\eta| < 0.5$.  Events are required to have a primary vertex position along the beam axis within 25~cm of the center of the TPC.  Tracks are required to have $p_{\text{T}} > 0.2~\text{GeV}/c$, at least 20 points measured in the TPC (out of a maximum of 45), a distance of closest approach to the collision vertex of less than 1~cm, and $|\eta| < 1$.  Events containing tracks with $p_{\text{T}} > 30~\text{GeV}/c$ are not considered because of poor momentum resolution.  Particle distributions are corrected for single particle tracking efficiency and for detector pair acceptance by event mixing (in relative azimuthal angle $\Delta\phi$ only).  

Jets are reconstructed from charged tracks in the TPC and neutral towers in the BEMC using the anti-$k_{\text{T}}$ algorithm~\cite{antikt} from the FastJet package~\cite{fastjet,fastjet3} with a resolution parameter $R = 0.4$. Only tracks with $p_{\text{T}} > 2~\text{GeV}/c$~ and towers with $E_{\text{T}} > 2~\text{GeV}$ are used in the jet reconstruction in order to control the effects of background fluctuations.  The reconstructed jet axis is required to be within $|\eta| < 1-R$.   The reconstructed trigger jet is the highest-$p_\text{T}$ jet that includes a BEMC tower that fired the HT trigger.  While in most jet reconstruction analyses it is necessary to subtract an average background energy from the reconstructed jet $p_{\text{T}}$~\cite{fastjetrhoA}, the $2~\text{GeV}$ cut on tracks and towers reduces the heavy-ion background significantly and makes a simple unfolding procedure more appropriate. 

\begin{figure}[!t]
\includegraphics[width=\linewidth]{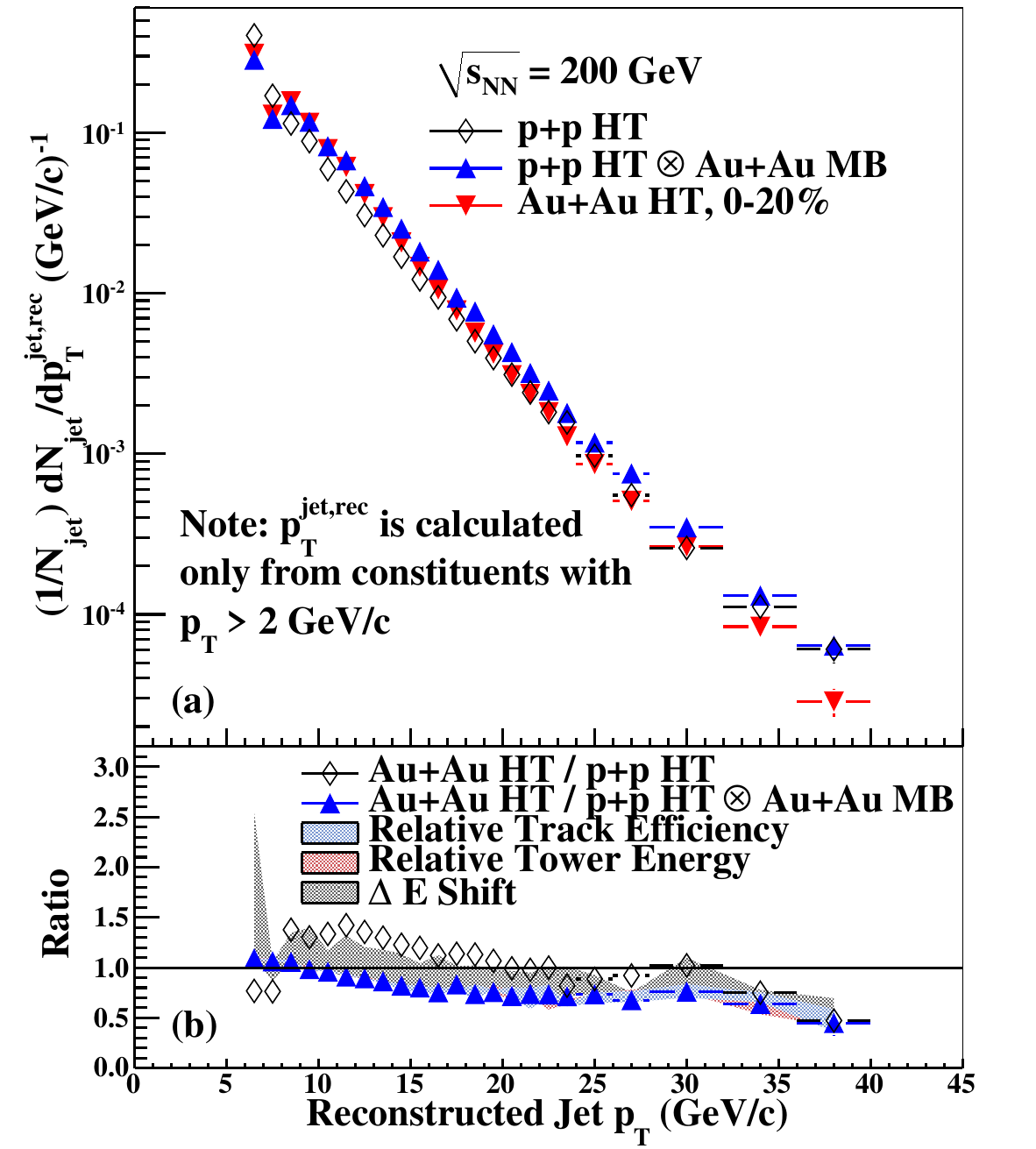}
\caption{\label{fig:jetPt}(Color online.)  (a) Detector-level $p_{\text{T}}^{\text{jet,rec}}$ spectra of HT trigger jets in $p$+$p$ and Au+Au, and of $p$+$p$ HT trigger jets embedded in Au+Au MB events.  (b) Ratio of $(1/N) dN/dp_{\text{T}}^{\text{jet,rec,Au+Au}}$ to $(1/N) dN/dp_{\text{T}}^{\text{jet,rec},p+p\text{~emb}}$ with uncertainties due to the relative tracking efficiency, relative tower energy, and $\Delta E = +1~\text{GeV}/c$ shift.  The ratio of $(1/N) dN/dp_{\text{T}}^{\text{jet,rec,Au+Au}}$ to $(1/N) dN/dp_{\text{T}}^{\text{jet,rec},p+p}$ is also shown.}
\end{figure}

In order to make quantitative comparisons between jets in Au+Au and $p$+$p$, it is necessary to compare jets with similar energies.  It is expected that the combination of the constituent $p_{\text{T}}$ cut and the HT trigger requirement biases the Au+Au jet population towards unmodified ($p$+$p$-like) jets~\cite{renk_bias}.  While the reconstructed jet $p_{\text{T}}$ is not directly related to the original parton energy, detector-level jets in Au+Au with a given $p_{\text{T}}^{\text{jet,rec,Au+Au}}$ are matched to similar detector-level $p$+$p$ jets using a bin-by-bin unfolding procedure.  The effect of the background associated with heavy-ion collisions on the trigger jet energy is assessed through embedding $p$+$p$ HT events in Au+Au minimum bias (MB) events (with the same high-multiplicity bias as the Au+Au HT events).  Under the assumption that Au+Au HT trigger jets are similar to $p$+$p$ HT trigger jets in a Au+Au collision background, the correspondence between the $p$+$p$ jet energy ($p_{\text{T}}^{\text{jet,rec},p+p}$) and the Au+Au jet energy ($p_{\text{T}}^{\text{jet,rec},p+p\text{~emb}} \simeq p_{\text{T}}^{\text{jet,rec,Au+Au}}$) can be determined through this embedding.  For a given range in $p_{\text{T}}^{\text{jet,rec},p+p}$ the corresponding $p_{\text{T}}^{\text{jet,rec,p+p\text{~emb}}}$ distribution is obtained.  When comparing Au+Au jets to equivalent $p$+$p$ jets in this analysis, the Au+Au signal is weighted according to this distribution.  This procedure largely accounts for the effects of background fluctuations in Au+Au events, as demonstrated in Fig.~\ref{fig:jetPt}(a)+(b).  Particularly at low $p_{\text{T}}$, the ratio of the $p_{\text{T}}^{\text{jet,rec,Au+Au}}$ spectrum to the $p_{\text{T}}^{\text{jet,rec},p+p}$ spectrum is restored to unity after embedding.  The possibility of additional discrepancies between the reconstructed jet energies in Au+Au and $p$+$p$, due to physics or other measurement effects, is included within systematic uncertainties.  

The performance of the TPC and BEMC can vary in different collision systems and over time.  These variations are accounted for in the relative tracking efficiency between Au+Au and $p$+$p$ ($90\% \pm 7\%$ for $p_{\text{T}} > 2~\text{GeV}/c$), the relative tower efficiency ($98\% \pm 2\%$), and the relative tower energy scale ($100\% \pm 2\%$).   These variations in detector performance were included, and their systematic uncertainties were assessed, in the $p$+$p$ HT $\otimes$ Au+Au MB embedding.  The effects of the relative tracking efficiency uncertainty and the tower energy scale uncertainty on the $p_{\text{T}}^{\text{jet,rec}}$ spectrum are shown in Fig.~\ref{fig:jetPt}(b).  The embedding also accounted for jet $v_2$ and its associated uncertainty.  The effects of the tower efficiency and jet $v_2$ on the jet energy scale are found to be negligible, as is the effect of varying the hadronic correction scheme on the final results.

\begin{figure}[!t]
\includegraphics[width=\linewidth]{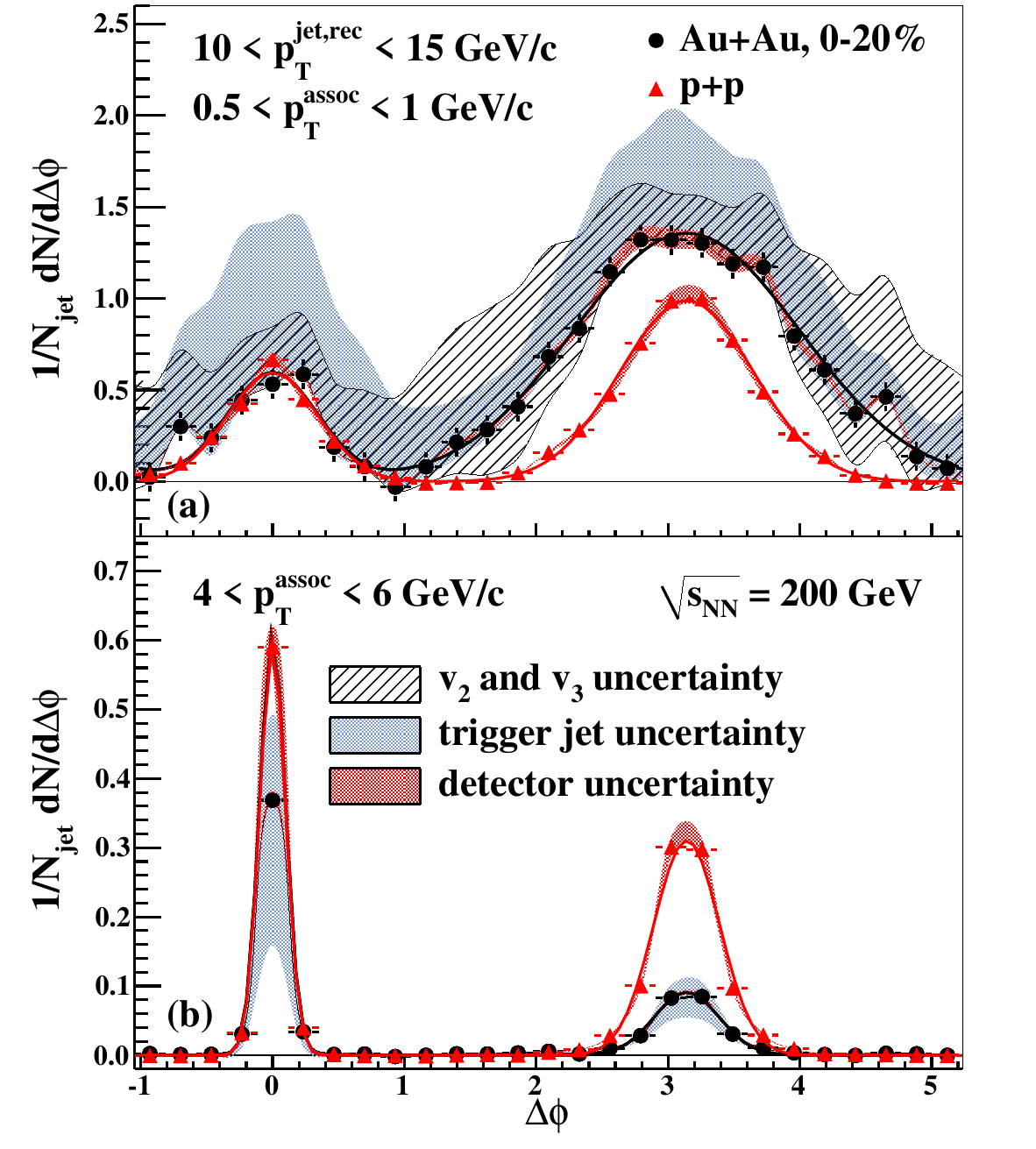}
\caption{\label{fig:raw}(Color online.)  Jet-hadron correlations after background subtraction for $10 < p_{\text{T}}^{\text{jet,rec}} < 15~\text{GeV}/c$ and for two ranges in $p_{\text{T}}^{\text{assoc}}$: (a) $0.5 < p_{\text{T}}^{\text{assoc}} < 1~\text{GeV}/c$ and (b) $4 < p_{\text{T}}^{\text{assoc}} < 6~\text{GeV}/c$.  The data points from Au+Au and $p$+$p$ collisions are shown with Gaussian fits to the jet peaks and systematic uncertainty bands due to: tracking efficiency, the shape of the combinatoric background, and the trigger jet energy scale.}
\end{figure}

Jet-hadron correlations are defined as distributions in $\Delta\phi = \phi_{\text{jet}} - \phi_{\text{assoc}}$, where $\phi_{\text{jet}}$ denotes the azimuthal angle of the axis of a reconstructed (trigger) jet and the associated particles are all charged hadrons, measured as TPC tracks, in the event.  To obtain the associated particle yields ($Y$) and widths ($\sigma$) of the jet peaks, the correlation functions are fit with the functional form: 
\begin{linenomath}\begin{align}\label{eq:bkgfit}
&\frac{Y_{\text{NS}}}{\sqrt{2\pi \sigma_{\text{NS}}^2}} e^{-(\Delta\phi)^2/2\sigma^2_{\text{NS}}} + \frac{Y_{\text{AS}}}{\sqrt{2\pi \sigma_{\text{AS}}^2}} e^{-(\Delta\phi-\pi)^2/2\sigma^2_{\text{AS}}}\\ \nonumber
&+ B\left(1+2v_2^{\text{assoc}}v_2^{\text{jet}}\cos(2\Delta\phi)+2v_3^{\text{assoc}}v_3^{\text{jet}}\cos(3\Delta\phi)\right),
\end{align}\end{linenomath}
which includes two Gaussians representing the trigger/nearside (NS) and recoil/awayside (AS) jet peaks, and a background term modulated by $v_2^{\text{assoc}}v_2^{\text{jet}}$ and $v_3^{\text{assoc}}v_3^{\text{jet}}$. Example $\Delta\phi$ correlations are shown in Fig.~\ref{fig:raw}, after the background term has been subtracted as detailed below.  

The elliptic anisotropy of the background is assumed to factorize into the product of the single-particle anisotropy of the associated particles due to elliptic flow ($v_2^{\text{assoc}}$) and the correlation of the jet axis with the $2^{\text{nd}}$-harmonic event plane ($v_2^{\text{jet}}$)~\cite{standardEP}.  The possibility that there is a correlation between the jet axis and the $3^{\text{rd}}$-harmonic event plane (which can give rise to a nonzero $v_3^{\text{assoc}}v_3^{\text{jet}}$ term) is also taken into account~\cite{alver}.  

The Gaussian yields of the jet peaks, $Y$, are integrated over a given bin in the transverse momentum of the associated hadrons ($p_{\text{T}}^{\text{assoc}}$), and the reconstructed jet $p_{\text{T}}$ ($p_{\text{T}}^{\text{jet,rec}}$), as well as over the $\Delta \eta$ acceptance.  

The effects of medium-induced modification can be quantified by the widths of the jet peaks, $\sigma$, as well as $D_{\text{AA}}$ and $\Sigma D_{\text{AA}}$, defined in Eqns.~(\ref{eq:Daa}) and~(\ref{eq:DeltaB}).  $D_{\text{AA}}$ measures the transverse-momentum difference between Au+Au and $p$+$p$ (in a given $p_{\text{T}}^{\text{assoc}}$ bin with mean $\langle p_{\text{T}}^{\text{assoc}}\rangle$): 
\begin{linenomath}\begin{align}\label{eq:Daa}
D_{\text{AA}}(p_{\text{T}}^{\text{assoc}}) \equiv &Y_{\text{Au+Au}}(p_{\text{T}}^{\text{assoc}}) \cdot \langle p_{\text{T}}^{\text{assoc}}\rangle_{\text{Au+Au}} \\ \nonumber
- &Y_{p+p}(p_{\text{T}}^{\text{assoc}}) \cdot \langle p_{\text{T}}^{\text{assoc}}\rangle _{p+p}. 
\end{align}\end{linenomath}
$\Sigma D_{\text{AA}}$ measures the energy balance over the entire $p_{\text{T}}^{\text{assoc}}$ range: 
\begin{linenomath}\begin{equation}\label{eq:DeltaB}
\Sigma D_{\text{AA}} \equiv \sum_{p_{\text{T}}^{\text{assoc}}\text{ bins}} D_{\text{AA}}(p_{\text{T}}^{\text{assoc}}). 
\end{equation}\end{linenomath}
If jets in Au+Au and $p$+$p$ have identical fragmentation patterns, then $D_{\text{AA}} = 0$ for all $p_{\text{T}}^{\text{assoc}}$.  Deviations from $D_{\text{AA}} = 0$ are indicative of jet modification.  

In order to analyze the jet correlation signal in Au+Au collisions it is necessary to subtract the large combinatoric background in heavy-ion collisions.  The background levels are estimated by fitting the functional form in~(\ref{eq:bkgfit}) to the $\Delta\phi$ distributions in Au+Au and $p$+$p$, with the flow terms constrained to zero in the latter. The shape of the Au+Au background is not well-constrained because $v_2^{\text{jet}}$ and $v_3^{\text{jet}}$ have not yet been measured experimentally (for the jet definition used in this analysis).  Therefore the uncertainties are investigated using two diametrically opposed assumptions.  To assess the effect of the uncertainty in the shape of the background, the assumption is made that Au+Au HT trigger jets undergo no medium modification.  Then to assess the effect of the uncertainty in the jet energy scale, the assumption is made that Au+Au HT trigger jets are maximally modified as described below.

First, it is assumed that Au+Au HT trigger jets undergo no modification and are equivalent to $p$+$p$ HT trigger jets (at all $p_{\text{T}}^{\text{assoc}}$).  When fitting the $\Delta\phi$ distributions with the functional form in~(\ref{eq:bkgfit}) the nearside yields and widths in Au+Au are fixed to the values measured in $p$+$p$, $v_2^{\text{assoc}}v_2^{\text{jet}}$ is fixed to a mean value and $v_3^{\text{assoc}}v_3^{\text{jet}}$ is left as a free parameter.  The mean $v_2^{\text{assoc}}$ is estimated to be the average of $v_2\text{\{FTPC\}}(p_{\text{T}}^{\text{assoc}})$ and $v_2\text{\{4\}}(p_{\text{T}}^{\text{assoc}})$, while $v_2^{\text{jet}}$ is estimated to be $v_2\text{\{FTPC\}}(6~\text{GeV}/c)$, where $v_2\text{\{FTPC\}}(p_{\text{T}})$ and $v_2\text{\{4\}}(p_{\text{T}})$ are parameterized from MB data in~\cite{STARv2}.  Here, $v_2\text{\{FTPC\}}$ is estimated with respect to the event plane determined in the Forward Time Projection Chambers ($2.4 < |\eta| < 4.2$)~\cite{STARftpc} and $v_2\text{\{4\}}$ is determined using the 4-particle cumulant method~\cite{cumulants}.  The $v_3^{\text{assoc}}v_3^{\text{jet}}$ values that result from the fits are reasonable compared to the data in~\cite{PHENIXv3,STARv3}.  The systematic uncertainties are determined by fixing $v_2^{\text{assoc}}v_2^{\text{jet}}$ to maximum and minimum values while letting $v_3^{\text{assoc}}v_3^{\text{jet}}$ float to force the Au+Au nearside yields to match $p$+$p$.  The limits on $v_2^{\text{assoc}}$ are estimated to be $v_2\text{\{4\}}(p_{\text{T}}^{\text{assoc}})$ and $v_2\text{\{FTPC\}}(p_{\text{T}}^{\text{assoc}})$. The bounds on $v_2^{\text{jet}}$ are conservatively estimated to be 70\% and 130\% of $v_2\text{\{FTPC\}}(6~\text{GeV}/c)$.  Additionally, it is observed in Fig.~\ref{fig:jetPt}(a) that the shape of the jet energy spectrum in Au+Au does not quite match the spectrum of $p$+$p$ HT jets embedded in Au+Au MB events.  The spectrum shape mismatch is covered by a $\Delta E = +1~\text{GeV}/c$ systematic uncertainty in the Au+Au trigger jet $p_{\text{T}}$, as shown in Fig.~\ref{fig:jetPt}(b).  

The second assumption is that the Au+Au HT trigger jets are maximally modified compared to $p$+$p$ HT trigger jets.  The background conditions that allow maximum increases in the nearside widths and yields are $v_2^{\text{assoc}}v_2^{\text{jet}} = 0$ and $v_3^{\text{assoc}}v_3^{\text{jet}} = 0$.  Under this assumption, the nearside $\Sigma D_{\text{AA}} = 0$ when the parent parton energies are correctly matched, even though $p_{\text{T}}^{\text{jet,rec,Au+Au}} \neq p_{\text{T}}^{\text{jet,rec},p+p}$ because $p_{\text{T}}^{\text{jet,rec}}$ is calculated only from tracks and towers above $2~\text{GeV}/c$.  The shift in the Au+Au trigger jet energy necessary to force $\Sigma D_{\text{AA}}$ to zero defines another systematic uncertainty estimate.  

\begin{figure}[!t]
\includegraphics[width=\linewidth]{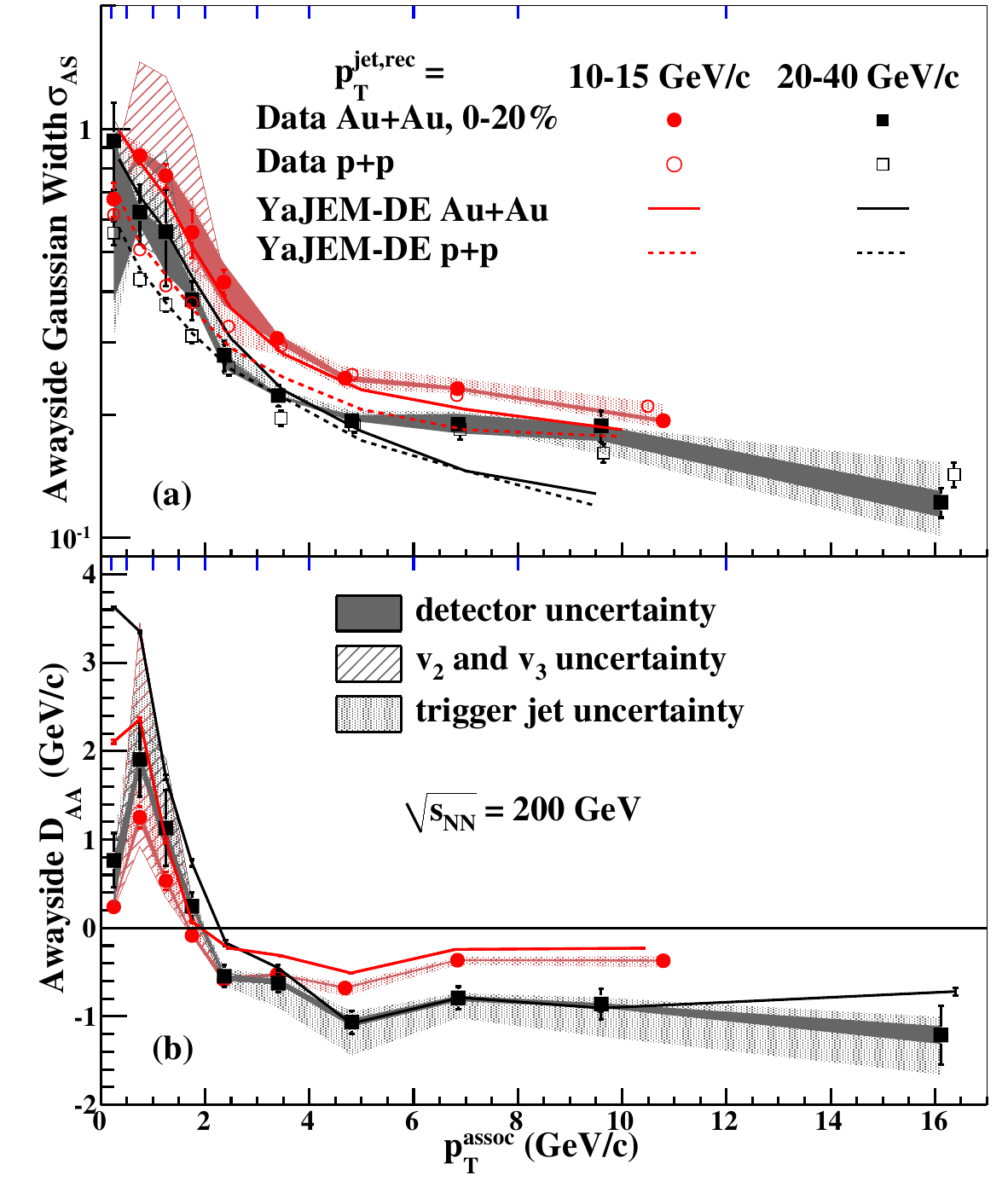}
\caption{\label{fig:c34}(Color online.)  The (a) Gaussian widths of the awayside jet peaks ($\sigma_{\text{AS}}$) in Au+Au (solid symbols) and $p$+$p$ (open symbols) and (b) awayside momentum difference $D_{\text{AA}}$ are shown for two ranges of $p_{\text{T}}^{\text{jet,rec}}$: $10-15~\text{GeV}/c$ (red circles) and $20-40~\text{GeV}/c$ (black squares).  Results for $15-20~\text{GeV}/c$ (not shown) are similar.  The boundaries of the $p_{\text{T}}^{\text{assoc}}$ bins are shown along the upper axes.  YaJEM-DE model calculations (solid and dashed lines) are from~\cite{yajemjh}.}
\end{figure}

The nearside jet is expected to have a surface bias~\cite{highpt_bias1,highpt_bias2,Renk_Dihadron} which makes it more likely that the recoil parton will travel a significant distance through the medium~\cite{surfacebias}, therefore enhancing awayside partonic energy loss effects.  
%The awayside Gaussian widths ($\sigma_{\text{AS}}$), $D_{\text{AA}}$, and $\Sigma D_{\text{AA}}$ are reported for different ranges in $p_{\text{T}}^{\text{jet,rec}}$. 
The awayside widths, shown in Fig.~\ref{fig:c34}(a), at high $p_{\text{T}}^{\text{assoc}}$ are the same in $p$+$p$ and Au+Au on average, indicating that jets containing high-$p_{\text{T}}$ fragments are not largely deflected by the presence of the medium.  The widths at low $p_{\text{T}}^{\text{assoc}}$ are indicative of broadening.  However, as the low-$p_{\text{T}}^{\text{assoc}}$ widths are anticorrelated with the magnitude of $v_3^{\text{assoc}}v_3^{\text{jet}}$, measurements of $v_n^{\text{jet}}$ are necessary before quantitative conclusions are drawn.  The awayside $D_{\text{AA}}$, shown in Fig.~\ref{fig:c34}(b), exhibits suppression of high-$p_{\text{T}}^{\text{assoc}}$ hadrons and enhancement of low-$p_{\text{T}}^{\text{assoc}}$ jet fragments in Au+Au, indicating that jets in Au+Au are significantly softer than those in $p$+$p$ collisions.  The amount of high-$p_{\text{T}}^{\text{assoc}}$ suppression, quantified by summing $D_{\text{AA}}$ only over bins with $p_{\text{T}}^{\text{assoc}} > 2~\text{GeV}/c$, ranges from $-2.5$ to $-5~\text{GeV}/c$ as jet $p_{\text{T}}$ increases.  Summing $D_{\text{AA}}$ over all $p_{\text{T}}^{\text{assoc}}$ bins to obtain the $\Sigma D_{\text{AA}}$ values, shown in Table~\ref{tab:DeltaB}, indicates that the high-$p_{\text{T}}^{\text{assoc}}$ suppression is balanced in large part by the low-$p_{\text{T}}^{\text{assoc}}$ enhancement.  

%\begin{figure}[!t]
%\includegraphics[width=\linewidth]{fig3.pdf}
%\caption{\label{fig:asw}(Color online.)  Gaussian widths of the awayside jet peaks ($\sigma_{\text{AS}}$) in Au+Au (solid symbols) and $p$+$p$ (open symbols) for two ranges of $p_{\text{T}}^{\text{jet,rec}}$: $10-15~\text{GeV}/c$ (red circles) and $20-40~\text{GeV}/c$ (black squares).  Results for $15-20~\text{GeV}/c$ (not shown) are similar.  The boundaries of the $p_{\text{T}}^{\text{assoc}}$ bins are shown along the upper axis.  YaJEM-DE model calculations (solid and dashed lines) are from~\cite{yajemjh}.}
%\end{figure}

Theoretical calculations from YaJEM-DE~\cite{yajemde}, a Monte Carlo model of in-medium shower evolution, are also shown for $\sigma_{\text{AS}}$ and $D_{\text{AA}}$ in Fig.~\ref{fig:c34}~\cite{yajemjh}.  This model incorporates radiative and elastic energy loss, and describes many high-$p_{\text{T}}$ observables from RHIC.  After the intrinsic transverse momentum imbalance, $k_{\text{T}}$, of the initial hard scattering was tuned to provide the best fit to the $p$+$p$ yields ($Y_{AS,p+p}$), this model largely reproduces several of the quantitative and qualitative features observed in data.  At high $p_{\text{T}}^{\text{assoc}}$ the Au+Au and $p$+$p$ widths match and the jet yields are suppressed, while the missing energy appears as an enhancement and broadening of the soft jet fragments.  

%\begin{figure}[!t]
%\includegraphics[width=\linewidth]{fig4.pdf}
%\caption{\label{fig:asd}(Color online.)  Awayside momentum difference $D_{\text{AA}}$ for two ranges of $p_{\text{T}}^{\text{jet,rec}}$: $10-15~\text{GeV}/c$ (red circles) and $20-40~\text{GeV}/c$ (black squares).  Results for $15-20~\text{GeV}/c$ (not shown) are similar.  The boundaries of the $p_{\text{T}}^{\text{assoc}}$ bins are shown along the upper axis.  YaJEM-DE model calculations (solid lines) are from~\cite{yajemjh}.}
%\end{figure}

To conclude, jet-hadron correlations are used to investigate the properties of the Quark-Gluon Plasma created in heavy-ion collisions by studying jet quenching effects.  The trigger/nearside jet sample is highly biased towards jets that have not interacted with the medium, which may enhance the effects of jet-quenching on the recoil/awayside jet.  While the widths of the awayside jet peaks are suggestive of medium-induced broadening, they are highly dependent on the shape of the subtracted background.  It is observed that the suppression of the high-$p_{\text{T}}$ associated particle yield is in large part balanced by low-$p_{\text{T}}^{\text{assoc}}$ enhancement.  The experimentally-observed redistribution of energy from high-$p_{\text{T}}$ fragments to low-$p_{\text{T}}$ fragments that remain correlated with the jet axis is consistent with radiative/collisional energy loss models for parton interactions within the Quark-Gluon Plasma.  

\begin{table}[!b]
\begin{tabular}{|c|c|c|c|c|}
\hline
$p_{\text{T}}^{\text{jet,rec}}$ & $\Sigma D_{\text{AA}}$ & Detector & $v_2$ and $v_3$ & Jet Energy Scale\\ 
($\text{GeV}/c$) & ($\text{GeV}/c$) & Uncert. & Uncert. & Uncert. \\ \hline
$10-15$ & $-0.6 \pm 0.2$ & $^{+0.2}_{-0.2}$ & $^{+3.7}_{-0.5}$ & $^{+2.3}_{-0.0}$\\ \hline
$15-20$ & $-1.8 \pm 0.3$ & $^{+0.3}_{-0.3}$ & $^{+1.0}_{-0.0}$ & $^{+1.9}_{-0.0}$\\ \hline
$20-40$ & $-1.0 \pm 0.8$ & $^{+0.1}_{-0.8}$ & $^{+1.2}_{-0.1}$ & $^{+0.3}_{-0.0}$\\ \hline
\end{tabular}
\caption{\label{tab:DeltaB} Awayside $\Sigma D_{\text{AA}}$ values with statistical and systematic uncertainties due to detector effects, the shape of the combinatoric background, and the trigger jet energy scale.}
\end{table}

\begin{acknowledgments}
We thank the RHIC Operations Group and RCF at BNL, the NERSC Center at LBNL, the KISTI Center in Korea and the Open Science Grid consortium for providing resources and support. This work was supported in part by the Offices of NP and HEP within the U.S. DOE Office of Science, the U.S. NSF, CNRS/IN2P3, FAPESP CNPq of Brazil, Ministry of Ed. and Sci. of the Russian Federation, NNSFC, CAS, MoST and MoE of China, the Korean Research Foundation, GA and MSMT of the Czech Republic, FIAS of Germany, DAE, DST, and CSIR of India, National Science Centre of Poland, National Research Foundation (NRF-2012004024), Ministry of Sci., Ed. and Sports of the Rep. of Croatia, and RosAtom of Russia.  Finally, we gratefully acknowledge a sponsored research grant for the 2006 run period from Renaissance Technologies Corporation.
\end{acknowledgments}

\bibliography{biblio}{}
\bibliographystyle{apsrev4-1}
\end{document}